\documentclass[preprint,prb,showpacs,superscriptaddress]{revtex4}
\usepackage{stmaryrd}
\usepackage{amsmath}
\usepackage{amssymb}
\usepackage{graphicx}
\usepackage{dcolumn}
\usepackage{bm}

\begin{document}

\begin{titlepage}
\title{Unconventional quantized edge transport in the presence of inter-edge coupling in intercalated graphene}

\author{Yuanchang Li\footnote{liyc@nanoctr.cn}}
\address{National Center for Nanoscience and Technology, Beijing 100190, People¡¯s Republic of China}
\date{\today}

\begin{abstract}
It is generally believed that the inter-edge coupling destroys the quantum spin Hall (QSH) effect along with the gap opening at the Dirac points. Using first-principles calculations, we find that the quantized edge transport persists in the presence of inter-edge coupling in Ta intercalated epitaxial graphene on SiC(0001), being a QSH insulator with the non-trivial gap of 81 meV. In this case, the band is characterized by two perfect Dirac cones with different Fermi velocities, yet only one maintains the edge state feature. We attribute such an anomalous behavior to the orbital-dependent decay of edge states into the bulk, which allows the inter-edge coupling just between one pair of edge states rather than two.
\end{abstract}

\maketitle
\draft

\vspace{2mm}

\end{titlepage}

\section{Introduction}

The groundbreaking work of the quantum Hall effect (QHE)\cite{QHE} has opened a new avenue for the studies of boundary state physics in condensed matter. Recently, a new member, called quantum spin Hall effect (QSHE)\cite{Kane,Bernevig}, is brought into the family of Hall systems. Different from the chiral edge states induced by the external magnetic field in the QHE, the QSHE occurs without need of the magnetic field and the strong spin-orbit coupling (SOC) itself leads to the time-reversal symmetry protected helical edge states. This yields a substantial difference from the QHE, i.e., the states on the opposite edges can couple with each other to generate a bandgap and destroy the QSHE as the system width is reduced to be rather narrow.\cite{ZhouB} Consequently, a topological phase transition occurs between QSH state and trivial insulator. Such an inter-edge coupling may also lead to rich physical phenomena as reported in other QH systems\cite{Chalker1,Chalker2,Chalker3}.

Due to the SOC, there are two pairs of Dirac states on the edges in a QSH insulator. In principle, the inter-edge coupling can be described by a 4 $\times$ 4 matrix with two parameters $R$ and $T$ related to the two characteristic tunneling processes\cite{Delplace}, as schematically illustrated in Fig. 1. Specifically, the process of $R$ ($T$) happens between the same (different) spins. $R$ and $T$ become significant only after a critical ribbon width. When the two edges are decoupled for a sufficient width, both $R$ and $T$ are zero and the system exists in the QSH state protected by the time-reversal symmetry.\cite{Wu} When the two edges are close enough to each other, neither $R$ nor $T$ is zero and the inter-edge coupling produces a gap in the spectrum, hence destroying the QSH effect\cite{ZhouB}. Intrinsically, the gap opening is resulted from the interactions of Dirac-fermions between two edges. However, the previous study\cite{us} revealed that the interactions of two Dirac cones can also preserve the perfect Dirac spectrum without opening a gap, just renormalizing the Fermi velocity if only one of $R$ and $T$ is zero. In other words, the Dirac spectrum would remain robust in the presence of inter-edge coupling. More importantly, Delplace et al.\cite{Delplace} have shown that the backscattering is prohibited too for this case and the system is reduced to two decoupled copies of quantum Hall edge states. So far, the knowledge corresponding to the two cases of $T=$ 0 and $R=$ 0 or $T\neq$ 0 and $R\neq$ 0 has been well-established\cite{ZhouB,Wu}. It is naturally to ask: Whether is there a system that can realize the case of just $T=$ 0 or $R=$ 0?

\begin{figure}[tbp]
\includegraphics[width=0.5\textwidth]{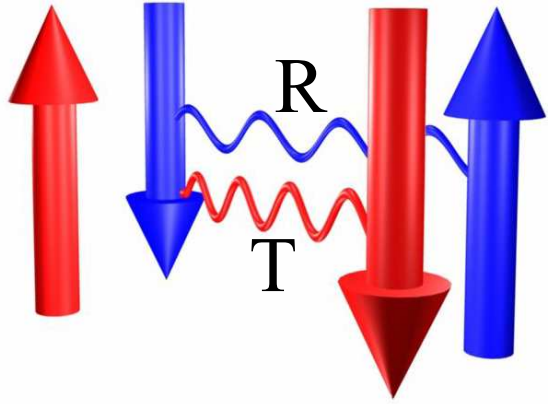}
\caption{\label{fig:fig1}  (Color online)  Illustration of the inter-edge coupling between the Dirac states on two edges. Red and blue represent the directions of spin while the arrows represent the directions of propagation. $R$ and $T$ are two interaction parameters corresponding to the two possible tunneling processes between the same and different spins. Generally speaking, there should exist three cases dependent upon the inter-edge coupling, i.e., (i) $R=$0 and $T=$0, (ii) $R\neq$0 and $T\neq$0, (iii) $R=$0 or $T=$0. The knowledge about the first two situations has been well-established but hitherto no report on the third case.}
\end{figure}

In this paper, we reveal that the Ta intercalated epitaxial graphene on SiC(0001) [denoted as G/\emph{i}-Ta/SiC] is such a system using the first-principles calculations. The large SOC of Ta opens a non-trivial gap of 81 meV, manifesting itself in the QSH state. A key difference from conventionally investigated QSH insulators is the significant Rashba splitting in G/\emph{i}-Ta/SiC, which leads to the distinct decay lengths of edge states into the bulk. Consequently, as the G/\emph{i}-Ta/SiC ribbon width reduces, three phases emerge in sequence, i.e., (I) the true time-reversal symmetry protected QSH state, (II) interacted quantized edge transport state and (III) trivial insulator state. In phase (II), there coexist the bulk and edge Dirac-fermions due to the inter-edge coupling only between one pair of edge states. Our findings are valuable to understand the topological phase transition and topologically protected quantized transport more deeply.

\begin{figure}[tbp]
\includegraphics[width=0.9\textwidth]{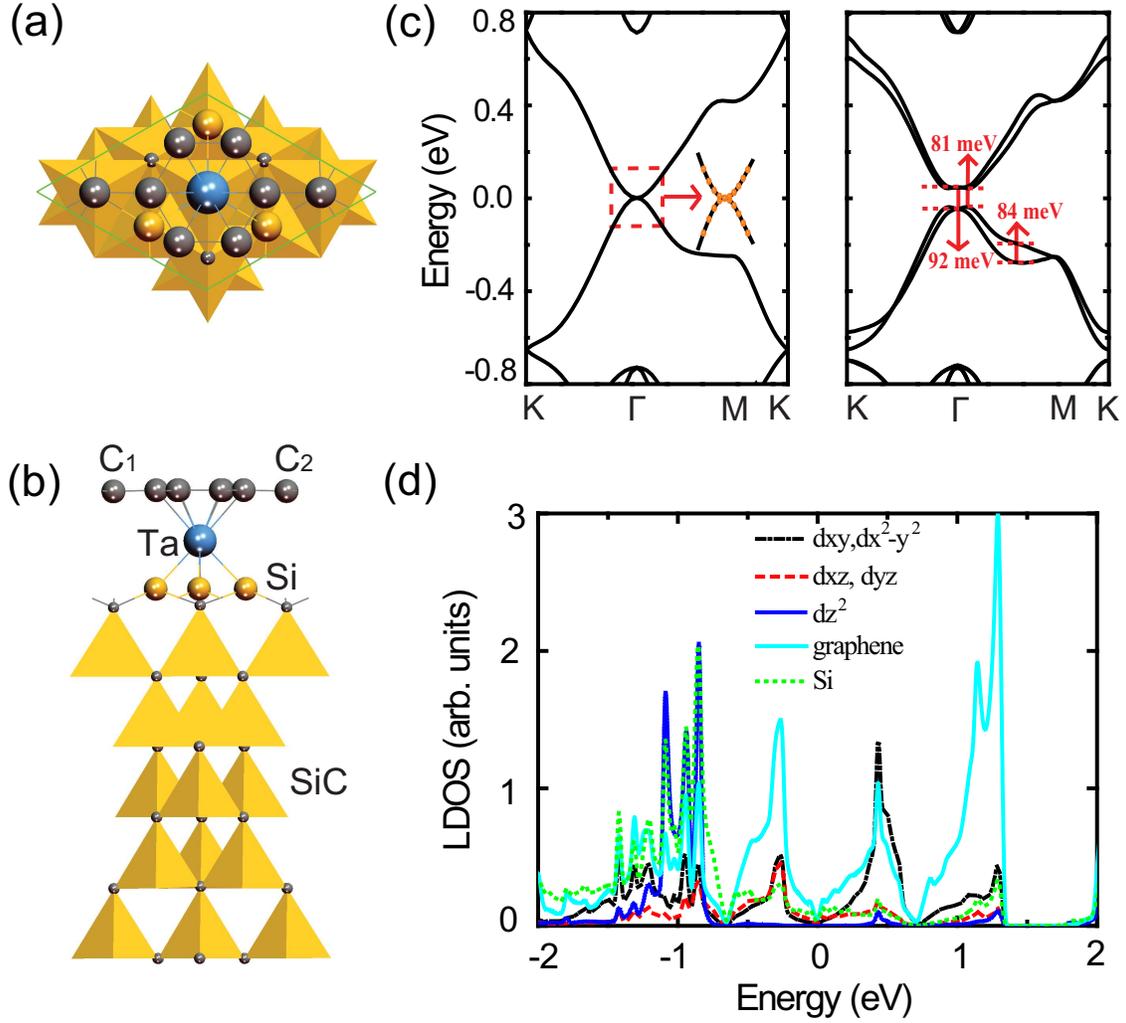}
\caption{\label{fig:fig2}  (Color online) Top (a) and side (b) view of the optimized geometries of the G/\emph{i}-Ta/SiC. Green dashed rhombus in (a) represents the surface cell. Note that C$_1$ and C$_2$ denote the two suspended carbon atoms not directly bonding to Ta. (c) Band structures of the G/\emph{i}-Ta/SiC system without (Left) and with (Right) spin-orbit coupling. Inset is the magnified plot of the parabolic dispersion around $\Gamma$ point. (d) Local density of states of G/\emph{i}-Ta/SiC system. The Fermi levels are set to zero.}
\end{figure}

\section{Method and models}

The calculations were performed using density-functional theory (DFT) with the projector augmented wave \cite{PAW} method and the local density approximation (LDA) \cite{CA} for the exchange and correlation potential, as implemented in the Vienna \emph{ab initio} simulation package \cite{vasp}. The cutoff energy was set to 400 eV. We place a 2 $\times$ 2 graphene overlayer on top of a $\sqrt3 \times \sqrt3R30^{\circ}$ 6$H$-SiC(0001) with one Ta atom intercalated between them (see Fig. 2). We model the SiC substrate with six SiC bilayers and fix the lower three at their respective bulk positions to simulate the bulk environment whereas fully relax all other atomic positions without any symmetry constraint until the residual forces are less than 0.01 eV/\AA. We choose the vacuum layer thickness larger than 10 \AA. In the calculations for very wide ribbons ($W$ = 26, 46), the SiC substrate is approximately modelled by one SiC bilayer, to achieve a balance between calculation efficiency and accuracy. Test calculations using narrower ribbons show that such a treatment yields an excellent description of the states near the Fermi level.

\section{Results and discussion}

We first explore the bulk properties of G/\emph{i}-Ta/SiC. Figure 2 plots its geometric configuration and the corresponding electronic structures without and with the SOC. It is worth emphasizing that many transition metal elements\cite{Gao,WangAPL,Eelbo,LiJACS,de Lima} have been successfully inserted into the interface between graphene overlayer and SiC substrate. In this configuration, all the dangling bonds of surface Si are fully saturated by the inserted Ta when the ratio of Ta to surface Si is 1/3, and the $p$-$d$ hybridization quenches the magnetic moment of transition metal, leaving the time-reversal symmetry reserved similar to the other transition metal intercalation\cite{us,TIus,usjpc}. When the SOC is not considered, two bands intersect each other just on the Fermi level [Left panel in Fig. 2(c)]. Different from that in graphene, the crossing point lies at $\Gamma$ point instead of $K$. Looked closely, it is found that the dispersion is largely parabolic in the region near $\Gamma$ point rather than linear as the magnified plot in the dashed rectangle. Taking into account the SOC, there opens a direct gap of 92 meV at $\Gamma$ point [Right panel in Fig. 2(c)].

Note that the system possesses a significant Rashba splitting, which lowers the system bandgap a little to 81 meV as shown in Fig. 2(c). Although it is still a direct gap semiconductor, the valence band top and conduction band bottom have a small displacement from $\Gamma$ point. The Rashba splitting is $k$-dependent strongly, ranging from zero to 84 meV for the topmost valence band. Such a large Rashba splitting is barely reported in the previous literatures, which may bring about new features to the helical edge states. Although the bulk band structure is not characterized by the linear Dirac spectrum as $d^5$ transition metal intercalated system\cite{TIus} and presents large Rashba splitting, the substantial SOC gap should also drive the system into a robust QSH state as previously demonstrated.\cite{HuJ}

We further calculate the local density of states (LDOS) as shown in Fig. 2(d). Owing to the local $C_{3v}$ symmetry, the degenerate Ta $d$-states split into three different subgroups: the $d_{z^2}$ singlet and the ($d_{xy}$, $d_{x^2-y^2}$) and ($d_{xz}$, $d_{yz}$) doublets. There appears considerable Ta $d$ states and surface Si $p$ states besides the graphene $p$ states around the Fermi level, implying their strong hybridization. This on the one hand endows the sandwiched structure a good stability and on the other hand, enhances the system SOC significantly, thereby the large SOC gap. Such a transition metal $d$-electrons dominant SOC gap has been well demonstrated by the previous studies\cite{HuJ,TIus} and is not the concern of this work.

In detail, the lowest unoccupied band is dominantly contributed from the Ta $d_{xy}$ and $d_{x^2-y^2}$ orbitals, whose hybridizations with C $p_z$ orbitals hold the quasi-2D inversion symmetry\cite{usprbRap}. Nevertheless, the ($d_{xy}$, $d_{x^2-y^2}$) and ($d_{xz}$, $d_{yz}$) doublets contribute more or less to the topmost occupied band. Note that the couplings of $d_{xz}$ and $d_{yz}$ with C $p_z$ orbitals break the quasi-2D inversion symmetry\cite{usprbRap}. This is probably why the Rashba splitting in valence band is more obvious compared with in conduction band [See the right panel of Fig. 2(c)].

\begin{figure}[tbp]
\includegraphics[width=0.85\textwidth]{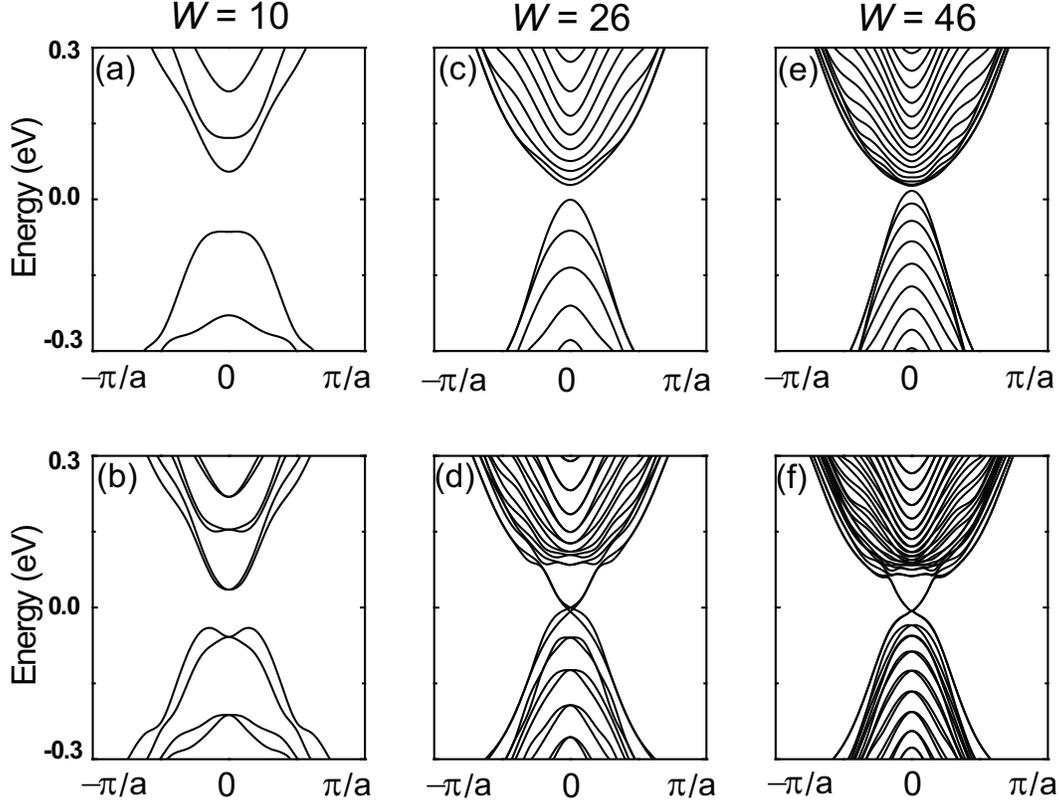}
\caption{\label{fig:fig3}  (Color online) (a) Band structures corresponding to different ribbon widths without (a)(c)(e) and with (b)(d)(f) spin-orbit coupling. The ribbons contain 10, 26 and 46 zigzag carbon dimers from left to right, respectively. Fermi levels are at energy zero.}
\end{figure}

The topologically non-trivial edge states are the direct evidence for the QSH insulator. Next we turn to study the electronic properties of G/\emph{i}-Ta/SiC ribbons with different widths. We consider the ribbons with zigzag termination because of not breaking the strong Ta-C and Ta-Si bonds (see Figs. 2(a) and (b)). Figure 3 shows the calculated electronic structures of zigzag G/\emph{i}-Ta/SiC ribbons for representative width ($W=10$, 26 and 46) without and with the SOC. Note that since the crossing occurs at the $\Gamma$ point for bulk band when the SOC is switched off, there are not the trivial gapless edge states even in the zigzag ribbons unlike in pristine graphene.

Not including the SOC, as expected, the quantum size effect opens a band gap at the $\Gamma$ point and the gap size decreases as the ribbon width increases (see the upper panel of Fig. 3). Switching on the SOC, the band structure becomes very interesting. For $W=10$ case, the SOC reduces the band gap at the $\Gamma$ point while the remarkable Rashba splitting between the two spins leads to the indirect band gap feature as shown in Fig. 3(b). Specifically, the Rashba splitting is rather asymmetric for the valence and conduction band. When the ribbon width increases to $W=26$, it is found that the SOC has closed the band gap induced by quantum size effect as reflected in Fig. 3(d). Now the Dirac cone spectrum has formed. (Noting that the Rashba SOC can mix the up and down spin and open a gap at the crossing point. Not only is this gap negligible herein but also it does not change the essentially topological nature.\cite{IJMPB}) Although the metallic states emerge spanning the SOC gap as expected, the asymmetric splitting of the Dirac cones is surprisingly found for the first time. It can be seen that the upper part is nearly degenerate as we usually observe while the lower one splits obviously. We also find a small Dirac point separation of 8 meV between the two Dirac cones. Further increasing the ribbon width to $W=46$, the splitting gets negligible both at the $\Gamma$ point and between the topmost two valence bands as shown in Fig. 3(f), meaning the thoroughly inter-edge decoupling.

\begin{figure}[tbp]
\includegraphics[width=0.9\textwidth]{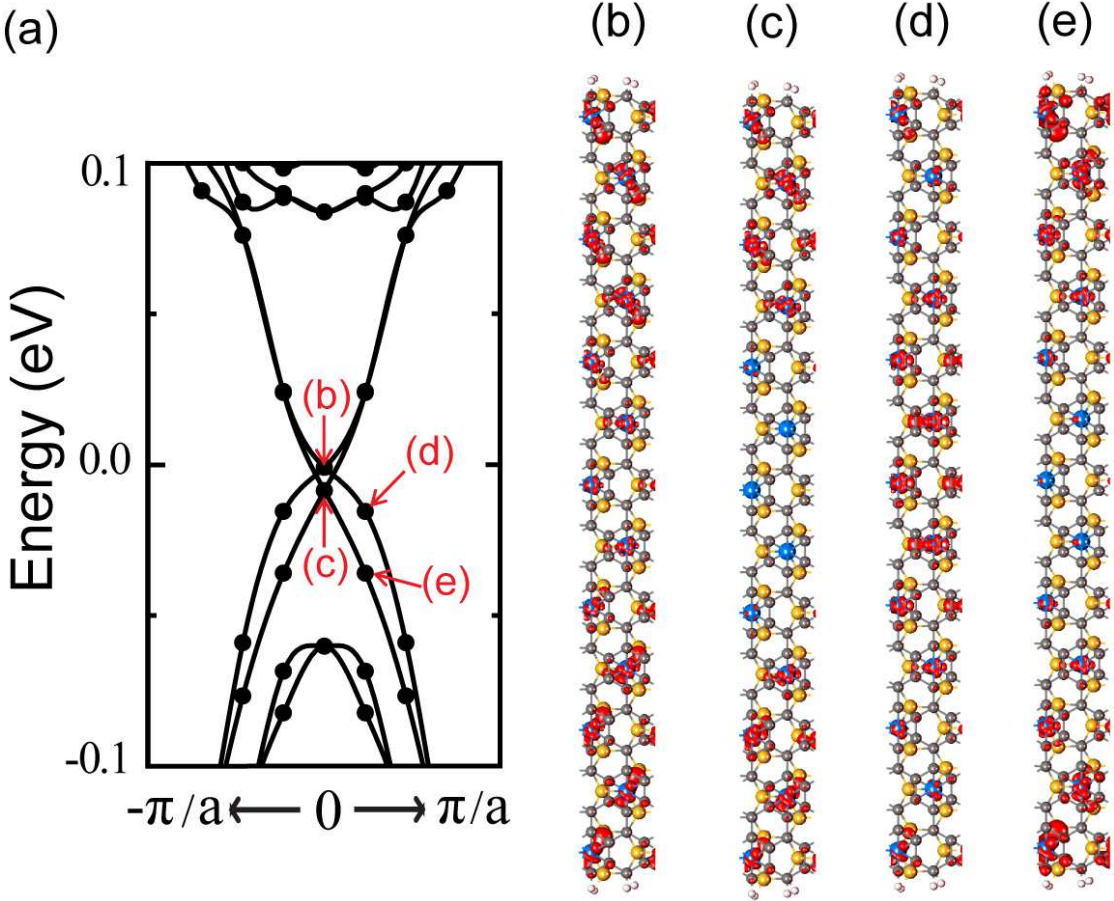}
\caption{\label{fig:fig4}  (Color online) (a) Electronic structure of zigzag G/\emph{i}-Ta/SiC ribbon with $W=26$ as well as the real-space charge distributions at the top (b) and down (c) Dirac point and the $k=$ 0.05 $\pi/c$ (c is the lattice constant) for the first (d) and second (e) topmost valence band with an isosurface = 0.0016 e/\AA$^3$. Note that only one of the two-fold degenerate states is plotted at the $\Gamma$ point.}
\end{figure}

Clearly, the strong inter-edge coupling destroys the QSH state in $W=10$ ribbon while the coupling is ignorable for $W=46$ one, allowing a true time-reversal symmetry protected QSHE. But the situation for $W=26$ ribbon becomes complicated. Apparently, the Dirac cone spectrum emerges, which would suggest the QSH state akin to the three dimensional counterparts\cite{Hek,ParkK}. However, our explored ribbons possess the mirror symmetry. Consequently, the opposite edge states have to be two-fold degenerate strictly and the energy degeneracy must be four at the $\Gamma$ point. Thus, the Dirac cone cannot be splitted as shown in Fig. 3(d) if there is no inter-edge coupling. These unambiguously show that $W=26$ ribbon is still too narrow to decouple the edge states on the opposite sides although the Dirac cone spectrum emerges. In addition, we find that the unique Dirac cone band structure is not related to the mirror symmetry of $W=26$ ribbon because a similar band is obtained in $W=28$ ribbon which does not possess the mirror symmetry.

To further explore the character of Dirac states in $W=26$ ribbon, we plot the real-space charge distribution at the represented $k$ points as shown in Fig. 4(a). It can be seen that the states shown in Figs. 4(b) and 4(d) distribute over the whole ribbon, meaning the bulk character. In sharp contrast, Figs. 4(c) and 4(e) both reveal the edge state character. Distinctly, there are two pairs of helical edge states for $W=46$ ribbon while they all exhibit the bulk character for $W=10$ ribbon. Indeed, there exists the inter-edge coupling for $W=26$ G/\emph{i}-Ta/SiC ribbon although it exhibits the unambiguous Dirac cone spectrum. These results are easily understood from the viewpoint of quantum confinement effect which promotes a width dependent gap and then this gap gets diminished upon increasing the values of $W$. An interesting finding here is the three physical phases associated with the inter-edge coupling, i.e., the survival or not of the edge states, or the hybridization of bulk and edge states.

In fact, the finite size effect is intrinsically the coupling between two Dirac cone states on the opposite edges. It may result in three kinds of band structures\cite{us}: (i) fully gapped Dirac cones due to two non-zero interaction parameters, (ii) two new Dirac cones with renormalized Fermi velocities under only one non-zero interaction parameter, (iii) intact Dirac cones due to two zero interaction parameters. The $W=10$, $W=26$ and $W=46$ ribbon exactly corresponds to the condition of (i), (ii) and (iii). One of the two Dirac cones belongs to the bulk state (See Figs. 4(b)-(d)) is a strong evidence of only one non-zero interaction parameter in $W=26$ case, i.e., either $R$ or $T$ is zero (See Fig. 1). According to Delplace et al.\cite{Delplace}, the system also prohibits the backscattering and hence the quantized edge transport. However, any disorder that changes the inter-edge coupling may destroy it. At this point, we realize a new topological phase associated with the band topology alone rather than time-reversal symmetry in $W=26$ G/\emph{i}-Ta/SiC ribbon.

Then a nature question arises: why there appears such an anomalous topological phase herein. As discussed above, the doublets ($d_{xy}$, $d_{x^2-y^2}$) and ($d_{xz}$, $d_{yz}$) have distinct effects on the quasi-2D inversion symmetry\cite{usprbRap} under the interaction with graphene $p_z$ orbitals, which leads to the coexistence of considerable intrinsic and Rashba SOC. Intuitively, the system could be considered to have two SOC gaps at different scales owing to the obvious Rashba splitting. This generally corresponds to the different SOC strengths, which are reported to be closely associated with the edge state decay length\cite{CanoPRL}. In fact, we can estimate the orbital-dependent decay length in quantity from the characteristic of topmost valence and lowest conduction band because of their distinct $d$-orbital contributions. For example, the lowest conduction band [dominated by ($d_{xy}$, $d_{x^2-y^2}$) doublet] is almost degenerate at $W=26$, meaning the fully decoupled edge states. So the decay length must be smaller than half the ribbon width, $\sim$3 nm. On the other hand, the splitting in the topmost valence band [dominated by both ($d_{xy}$, $d_{x^2-y^2}$) and ($d_{xz}$, $d_{yz}$) doublets] remains significant even if the ribbon width achieves $\sim$10 nm ($W=38$). Thus, the addition of $d_{xz}$ and $d_{yz}$ orbitals leads to a much larger decay length, $>$5 nm.

\begin{figure}[tbp]
\includegraphics[width=0.7\textwidth]{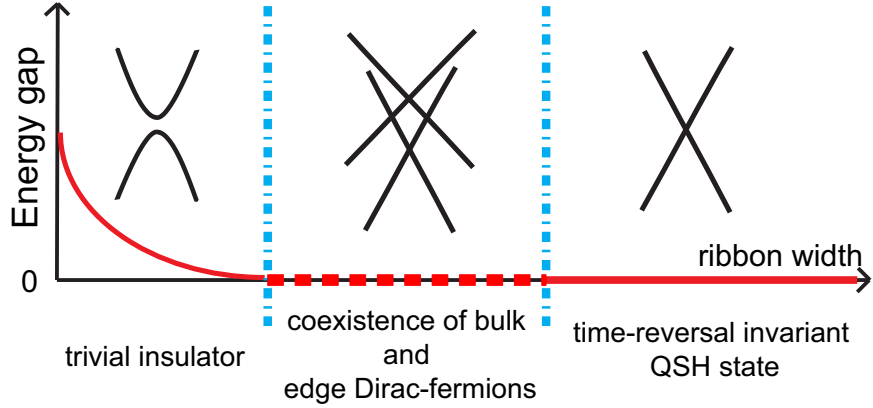}
\caption{\label{fig:fig5}  (Color online) Electronic properties of G/\emph{i}-Ta/SiC ribbon as a function of the width. When the ribbon is rather narrow, the Dirac cone is fully gapped due to the quantum size effect. As the increase of ribbon width, the gap decreases and is finally closed, meaning a topological phase transition. However, there still exists the inter-edge coupling and the Dirac-fermions are contributed both from bulk and edge. Further increasing the ribbon width, the inter-edge coupling becomes negligible and the system enters the true time-reversal invariant QSH state.}
\end{figure}

Figure 5 illustrates the general phase scheme of the QSH insulator as a function of its ribbon width. For the system with negligible Rashba SOC, there exists only two phases (Left and right panel in Fig. 5). The Dirac cone emerges exactly when the contributions from quantum size effect and the intrinsic SOC cancel each other and transition therein occurs between topologically trivial and non-trivial phases. However, if the system possesses a considerable Rashba SOC, the gap may close even though the contribution of intrinsic SOC is smaller than the quantum size effect. This will give rise of a new topological phase (Middle panel in Fig. 5) between the above two. Now the system can also exhibit quantized edge transport, but it substantially differs from the time-reversal protected QSH state. Such an intermediate state maintains until the quantum size effect induced gap is thoroughly closed by the intrinsic SOC, then entering the true time-reversal invariant QSH phase.

\section{Conclusion}

To summarize, we investigate the evaluation of edge states as the ribbon width in the QSH insulator of G/\emph{i}-Ta/SiC. We find a new topological phase that has not been noticed to date, locating between the well-developed true time-reversal invariant QSH phase and trivial insulator. In this new phase, the quantized edge transport is robust against the inter-edge coupling and no Dirac gap is opened. There is only one pair of edge states rather than two, distinctly different from the true time-reversal QSH state. We attribute this to the orbital-dependent decay of edge state into the bulk. Our work deepens the understanding of topological phase transition as well as the finite size effect in the QSH insulator.

\section{Acknowledgement}

The authors thank Damien West and S. B. Zhang for helpful discussions. We acknowledge the support of the National Natural Science Foundation of China (Grant Nos. 11304053).


\begin{references}
\bibitem{QHE} K. V. Klitzing, G. Dorda, and M. Pepper, Phys. Rev. Lett. \textbf{45}, 494 (1980).

\bibitem{Kane} C. L. Kane and E. J. Mele, Phys. Rev. Lett. \textbf{95}, 146802 (2005).

\bibitem{Bernevig} B. A. Bernevig, T. L. Hughes, and S. C. Zhang, Science \textbf{314}, 1757-1761 (2006).

\bibitem{ZhouB} B. Zhou, H. Z. Lu, R. L. Chu, S. Q. Shen, and Q. Niu, Phys. Rev. Lett. \textbf{101}, 246807 (2008).

\bibitem{Chalker1} J. T. Chalker and A. Dohmen, Phys. Rev. Lett. \textbf{75}, 4496 (1995).

\bibitem{Chalker2} J. J. Betouras and J. T. Chalker, Phys. Rev. B \textbf{62}, 10931 (2000).

\bibitem{Chalker3} J. T. Chalker, Yuval Gefen, and M. Y. Veillette, Phys. Rev. B \textbf{76}, 085320 (2007).

\bibitem{Delplace} P. Delplace, J. Li, and M. B\"{u}ttiker, Phys. Rev. Lett. \textbf{109}, 246803 (2012).

\bibitem{Wu} C. J. Wu, B. A. Bernevig, and S. C. Zhang, Phys. Rev. Lett. \textbf{96}, 106401 (2006).

\bibitem{us} Y. C. Li, P. C. Chen, G. Zhou, J. Li, J. Wu, B. -L. Gu, S. B. Zhang, and W. H. Duan, Phys. Rev. Lett. \textbf{109}, 206802 (2012).

\bibitem{PAW} P. E. Bl\"{o}chl, Phys. Rev. B \textbf{50}, 17953 (1994).

\bibitem{CA} D. M. Ceperley, and B. J. Alder, Phys. Rev. Lett. \textbf{45}, 566 (1980).

\bibitem{vasp} G. Kresse, and J. Furthm\"{u}ller, Phys. Rev. B \textbf{54}, 11169-11186 (1996).

\bibitem{Gao} T. Gao, Y. B. Gao, C. Z. Chang, Y. B. Chen, M. X. Liu, S. B. Xie, K. He, X. C. Ma, Y. F. Zhang, and Z. F. Liu, ACS nano \textbf{6}, 6562, (2012).

\bibitem{Eelbo} T. Eelbo, M. Wa\'{s}niowska, P. Thakur, M. Gyamfi, B. Sachs, T. O. Wehling, S. Forti, U. Starke, C. Tieg, A. I. Lichtenstein, and R. Wiesendanger, Phys. Rev. Lett. \textbf{110}, 136804 (2013).

\bibitem{WangAPL} Z. J. Wang, Y. Y. Dong, M. M. Wei, Q. Fu, and X. H. Bao, Appl. Phys. Lett. \textbf{104}, 181604, (2014).

\bibitem{de Lima} L. H. de Lima, R. Landers, and A. de Siervo, Chem. Mater. \textbf{26}, 4172, (2014).

\bibitem{LiJACS} G. Li, H. T. Zhou, L. D. Pan, Y. Zhang, L. Huang, W. Y. Xu, S. X. Du, M. Ouyang, A. C. Ferrari, and H. J. Gao, J. Am. Chem. Soc. \textbf{137}, 7099, (2015).

\bibitem{usjpc} Y. C. Li, J. Phys. Chem. C \textbf{120}, 2254, (2016).

\bibitem{TIus} Y. C. Li, P. Z. Tang, P. C. Chen, J. Wu, B. -L. Gu, Y. Fang, S. B. Zhang, and W. H. Duan, Phys. Rev. B \textbf{87}, 245127 (2013).

\bibitem{HuJ} J. Hu, J. Alicea, R. Q. Wu, and M. Franz, Phys. Rev. Lett. \textbf{109}, 266801 (2012).

\bibitem{usprbRap} Y. C. Li, D. West, H. Q. Huang, J. Li, S. B. Zhang, and W. H. Duan, Phys. Rev. B \textbf{92}, 201403(R) (2015).

\bibitem{IJMPB} C. L. Kane, Int. J. Mod. Phys. B \textbf{21}, 1155 (2007).

\bibitem{Hek} Y. Zhang, K. He, C. Z. Chang, C. L. Song, L. L. Wang, X. Chen, J. F. Jia, Z. Fang, X. Dai, W. Y. Shan, S. Q. Shen, Q. Niu, X. L. Qi, S. C. Zhang, X. C. Ma, and Q. K. Xue, Nature Phys. \textbf{6}, 584-588 (2010).

\bibitem{ParkK} K. Park, J. J. Heremans, V.W. Scarola, and Djordje Minic, Phys. Rev. Lett. \textbf{105}, 186801 (2010).

\bibitem{CanoPRL} L. Cano-Cort\'{e}s, C. Ortix, and J. van den Brink, Phys. Rev. Lett. \textbf{111}, 146801 (2013).
\end{references}
\end{document}